\documentclass[prb,twocolumn,showpacs,amsmath,amssymb
 reprint,
 amsmath,amssymb,
 aps,
prb,
]{revtex4-2}

\usepackage{graphicx}
\usepackage{dcolumn}
\usepackage{bm}
\usepackage{xcolor}
\usepackage{empheq}
\usepackage{amsmath}
\usepackage{mhchem}

\newcommand*\mybluebox[1]{%
  \colorbox{blue!10}{\hspace{1em}#1\hspace{1em}}}
  {\endEmphEqMainEnv}  

\newcommand*\myredbox[1]{%
  \colorbox{red!10}{\hspace{1em}#1\hspace{1em}}}
  {\endEmphEqMainEnv}  

  \newcommand*\mygreenbox[1]{%
  \colorbox{green!10}{\hspace{1em}#1\hspace{1em}}}
  {\endEmphEqMainEnv}  

\begin{document}

\preprint{APS/123-QED}

\title{Influence of electronic correlations on disorder-induced loop currents in high-Tc superconductors}

\author{Marius Paul and Götz Seibold}
\affiliation{
  Institut für Physik, BTU Cottbus-Senftenberg, PBox 101344, 03013 Cottbus, Germany
}%


\date{\today}

\begin{abstract}
Time-reversal symmetry breaking in superconducting systems can manifest itself in the form of currents which are induced by inhomogeneities in the charge and order parameter distribution. With regard to cuprates such states have been theoretically studied in the overdoped region of the phase diagram where the inhomogeneities are related to out-of plane dopants. In this paper we will extend previous work by including local correlations within the unrestricted Gutzwiller approximation in order to study its impact on the induced loop currents. In addition, we investigate the effect of next-nearest neighbor hopping which extends the TRSB phase towards half-filling. We find that in general correlations lead to a suppression of loop currents, however, the Gutzwiller approach can sustain such states to larger values of the local on-site repulsion as compared to the Hartree-Fock approximation. Our investigations allow for an estimation of the local magnetic moment emerging
from impurity-induced loop currents for cuprate superconductors.
\end{abstract}

\maketitle


\section{\label{sec:level1}Introduction}
The possible emergence of time reversal symmetry breaking (TRSB) superconducting states has attracted significant interest for decades in the field of unconventional superconductivity.
Basically, the appearance of nodes in the superconducting order parameter of these systems makes them susceptible to a phase change in order to create additional gapped regions \cite{sigrist98} which may trigger the formation of TRSB currents.
One prominent example for an unconventional TRSB SC is the case of Sr$_2$RuO$_4$
which for a long time has been believed to be a triplet SC with a chiral p-wave OP \cite{RevModPhys.75.657} until more recent
NMR-measurements \cite{Pustogow_2019, Ishida_2020, Chronister_2021} led to the conclusion, that the OP rather belongs to a spin-singlet state, even though there are still some unresolved issues \cite{maeno2024mysteryyearsunconventional}.
Ultrasound attenuation experiments on Sr$_2$RuO$_4$ \cite{Benhabib_2020, osti_1878571} have revealed, that the OP is of two-component-symmetry (i.e. two-dimensional irreducible representations of the $D_{4h}$ point group), whereas $\mu$SR(muon-spin relaxation)-measurements by \cite{Grinenko_2021} lead to the conclusion that the OP is chiral, i.e. that the two symmetry components have a relative phase of $\pi/2$. It is therefore a symmetry-protected TRSB state, which could also be confirmed by polar Kerr experiments \cite{PhysRevLett.97.167002}. 


In this paper we are investigating the possible
emergence of TRSB states in high-T$_c$ cuprates
which, in the clean case, are usually characterized by a d-wave SC order parameter. Therefore, the bulk of these systems does not break time reversal symmetry in the superconducting phase, as has been supported experimentally for YBa$_2$Cu$_3$O$_7$ (YBCO) and Bi$_2$Sr$_{2-x}$La$_x$CuO$_{6+\delta}$ by $\mu$SR-measurements \cite{PhysRevLett.64.2082, PhysRevB.88.180501, PhysRevB.75.054511}. $\mu$SR is able to detect small local magnetic fields, which are an important manifestation of TRSB in a macroscopic system. On the other hand, in the pseudogap phase, polarized neutron diffraction experiments found intra-unit cell magnetic order \cite{PhysRevLett.96.197001, PhysRevB.78.020506, PhysRevB.83.104504, Mangin-Thro2015, PhysRevLett.118.097003, Li2008, PhysRevB.84.224508, PhysRevB.98.214418, PhysRevLett.105.027004, PhysRevB.86.020504, PhysRevB.89.094523} in four different kinds of cuprates, which has been associated with loop currents in the CuO$_2$ planes, compatible with the original proposal by Varma \cite{varma97}, see also Ref. \onlinecite{CRPHYS_2021__22_S5_7_0}. Those findings have been supported further by experimental techniques utilising the Kerr effect \cite{PhysRevLett.100.127002}.

There are several mechanisms which lead to TRSB in the presence of disorder \cite{10.3389/fphy.2024.1353425}. E.g., non-magnetic-disorder-induced TRSB is predicted in form of local magnetic moments in disordered correlated models (Hubbard, tJ) including d-wave superconductivity \cite{PhysRevB.64.140501, PhysRevLett.89.217002, PhysRevLett.89.067003, PhysRevLett.92.077203, PhysRevLett.99.147002, PhysRevB.75.054520, PhysRevLett.105.147002, Schmid_2010, PhysRevB.88.220509, PhysRevLett.113.127001, Gastiasoro2015, PhysRevB.92.224510, PhysRevLett.117.257002, PhysRevB.99.014509, RevModPhys.81.45}. 
Furthermore, in their study
of the metal-to-insulator transition in overdoped
cuprates, Li et al. \cite{Kivelson} numerically found that
equilibrium loop currents occur in d-wave superconductors
in the presence of multiple (non-magnetic) impurities.
These currents could be traced back by Breiø et al. \cite{Clara_PhysRevB.105.014504} to locally formed TRSB $s+id$-regions, caused by the local change of the OP due to the impurities. Unlike the chiral state of Sr$_2$RuO$_4$, the $s+id$-state is a so-called accidental TRSB state, which means, that the two complex-combined symmetry components $s$ and $d$ are not fulfilling the symmetry relations set by a two-dimensional irreducible representation. Given a chiral state, as e.g. $p_x + i p_y$ and only one single impurity, TRSB can be present in form of one-directional loop currents flowing around the impurity \cite{PhysRevB.104.094505}, exhibiting a net current over the whole sample. But in the case of a non-chiral state, several clockwise and counter-clockwise loop currents are distributed around the impurity \cite{PhysRevLett.102.217002, PhysRevB.91.161102, PhysRevB.94.064519, PhysRevLett.116.097002}, as e.g. in the $s+id$ state. In the latter case, the sum of the vortices over the whole sample is zero. 

Previous studies on disorder induced TRSB in cuprates (for a review see \cite{10.3389/fphy.2024.1353425}) were based on the Bogoljubov-de Gennes (BdG) approach, thus neglecting the influence of strong correlations
in these materials. Here we want to bridge this gap and investigate the influence of electronic correlations on non-magnetic disorder-induced loop currents by means
of the unrestricted Gutzwiller approach \cite{sei98}
extended towards the inclusion of pairing correlations
\cite{GS_FB_08}.
In Sec. \ref{sec:form} we present the formalism underlying our investigations and discuss our results
in Sec. \ref{sec:results}. We first determine the phase diagram, including the $s+id$ phase for the homogeneous system in Sec. \ref{sec:hom}, then evaluate the single impurity case in Sec. \ref{sec:single} before the case for a concentration of impurities is studied in Sec. \ref{sec:multiple}.
We finally conclude our discussion in \ref{sec:sum}.


\

\section{Formalism}\label{sec:form}

Our investigations are based on the extended single-band Hubbard-hamiltonian supplemented with an intersite attractive interaction and local disorder

\begin{align}
    \hat{H} =& -\sum_{ i,j , \sigma} (t_{ij} + \mu \delta_{ij}) c^{\dagger}_{i\sigma}c_{j\sigma} +  U\sum_i \hat{n}_{i \uparrow} \hat{n}_{i \downarrow} \notag \\ 
    &+ \frac{V}{2} \sum_{\langle ij \rangle, \sigma, \sigma'} \hat{n}_{i\sigma}\hat{n}_{j\sigma'}  + \sum_{i, \sigma} V^{\text{imp}}_i \hat{n}_{i\sigma} \label{eq:ham}\;, 
\end{align}

where $t_{ij}=t$ for nearest neighbors, $t_{ij}=t'$ for nearest diagonal neighbors, and $\hat{n}_{i\sigma}=c_{i\sigma}^{\dagger} c_{i\sigma}$. For the nearest-neighbor interaction $\sim V<0$ we use the following mean-field decoupling

\begin{align*}
    \frac{V}{2} \sum_{\langle ij \rangle, \sigma, \sigma'} n_{i\sigma}n_{j\sigma'} \approx & \frac{V}{2} \sum_{\langle ij \rangle}  \Delta_{ij} \left(c^{\dagger}_{i\uparrow}c^{\dagger}_{j\downarrow} + c^{\dagger}_{j \uparrow}c^{\dagger}_{i \downarrow} \right)  \\
    &-  W_{ij} \left( c^{\dagger}_{i \uparrow} c_{j \uparrow} + c^{\dagger}_{i \downarrow} c_{j \downarrow}\right) + \text{h.c.} \\
    & - 2  |\Delta_{ij}|^2 + 2 |W_{ij}|^2 \;,
\end{align*}

where $\Delta_{ij} = \langle c_{j\downarrow} c_{i \uparrow} \rangle $ and $W_{ij} = \langle c^{\dagger}_{j \uparrow} c_{i \uparrow} \rangle$. Recall that for singlet-pairing states, only $s$- and $d$-wave symmetrical order parameters (OP) or a complex combination of those are possible, and we neglect the possibility of  $p$-wave-symmetry, which for charge concentration close to half-filling is suppressed for sufficient large values of $U/t$ \cite{cao2025pwavesuperconductivityinducednearestneighbor}. Throughout this work we set $V/t=-2$.  $V^{\text{imp}}_i$ is the impurity potential at site $i$. The local repulsion $\sim U$ is treated
within an unrestricted Gutzwiller approximation (GA)
\cite{sei98} which can be implemented 
by either a variational ansatz \cite{geb90} or the
Kotliar-Ruckenstein slave-boson scheme \cite{Kotliar_Ruckenstein}. Furthermore, in the presence
of superconductivity the unrestricted GA has to be applied in its charge-rotationally invariant form \cite{GS_FB_08} and one finds (cf. appendix \ref{sec:GAFORM}) that the kinetic energy is modified to 

\begin{align*}
    E_{\text{kin}} =& - \sum_{\langle ij \rangle, \sigma} t_{ij} z_i z_j \langle c^{\dagger}_{i \sigma}c_{j\sigma}\rangle \;,
\end{align*}
where the renormalization factors $z_i(\tilde{D},n_i,\Delta_i)$ depend on variational parameters $\tilde{D}_i$ corresponding to the double occupancy, the local
densities $n_i$ and local anomalous correlations
$\Delta_i=\langle c_{i\downarrow}c_{i\uparrow}\rangle$.
We also compare the GA results with those obtained from
a Hartree-Fock (HF) decoupling of the onsite-interaction

\begin{align*}
    U\sum_i n_{i, \uparrow} n_{i, \downarrow} \approx& U\sum_i\left\lbrack  \Delta_i c^{\dagger}_{i \uparrow}c^{\dagger}_{i \downarrow} + \text{h.c.}-|\Delta_i|^2 \right\rbrack \\
    +& \frac{U}{2}\sum_{i,\sigma} \left\lbrack n_i c_{i\sigma}^\dagger c_{i\sigma}-\frac{1}{4} n_i^2 \right\rbrack
    \,.
\end{align*}

In both methods (HF and GA) the effective single-particle hamiltonian $\hat{H}$ is solved iteratively within the BdG approach  on a $N\times N$ square lattice with lattice constant $a$ and periodic boundary conditions, using as convergence criterion at iteration step $n$ that $|C[n+1] - C[n]|/|C[n]|<10^{-6}$ with $C$ being the vector containing all expectation values $\Delta_{ij}$, $W_{ij}$, $\langle n_{i\uparrow} \rangle$ and $\Delta_i$. The chemical potential $\mu$ is adjusted to match the mean electron density $n_{\text{el}}$. The temperature is fixed at $k_B T = 10^{-3}t$. 
In case of GA, in addition to self-consistency, one has to minimize the total energy with respect to the double
occupancy variational parameters.
Note, that for $U/t=0$ both methods (HF and GA) lead to an identical ground state of the system. 
The emerging TRSB currents within the sample can be derived by combining Heisenberg's equation of motion $i \hbar \frac{\text{d}}{\text{d} t} \hat{n}_{i,\sigma}  =  [\hat{n}_{i,\sigma}, \hat{H}]$ with the continuity equation $i \hbar \frac{\text{d}}{\text{d} t} \hat{n}_{i,\sigma} + [div\vec{j}]_i=0$.
One obtains

\begin{align*}
    \langle j_{ij} \rangle_{\text{HF}} &= 4 \frac{e t_{ij}}{\hbar} \text{Im} (W_{ij}) \;, \\
    \langle j_{ij} \rangle_{\text{GA}} &= 4 \frac{e t_{ij} z_i z_j }{\hbar} \text{Im} (W_{ij}) \;,
\end{align*}

where we have defined the current $j_{ij}$ as moving electron charge $e$ from site $i$ to $j$. Details about the computational procedure can
be found in appendix \ref{sec:GAFORM}.




\section{Results}\label{sec:results}

Here we are concerned with TRSB from $s$-and $d$-wave
superconducting states which are accidentally degenerate and may form complex superpositions in order to minimize the free energy of the system \cite{10.3389/fphy.2024.1353425}. 
In Sec. \ref{sec:hom} we will first analyze the 
impact of 
local correlations on the stability of the resulting $s+id$ phase in the homogeneous system, similar to 
previous work \cite{mueller_prb14,roemer_prb92} which, however, is in the context of spin mediated superconductivity.  
Secs. \ref{sec:single}, \ref{sec:multiple} we investigate
the emerging currents in the presence of local impurities.

\subsection{Phase diagrams of the homogeneous system}\label{sec:hom}
\begin{figure}
     \centering
     \includegraphics[width=0.47\textwidth]{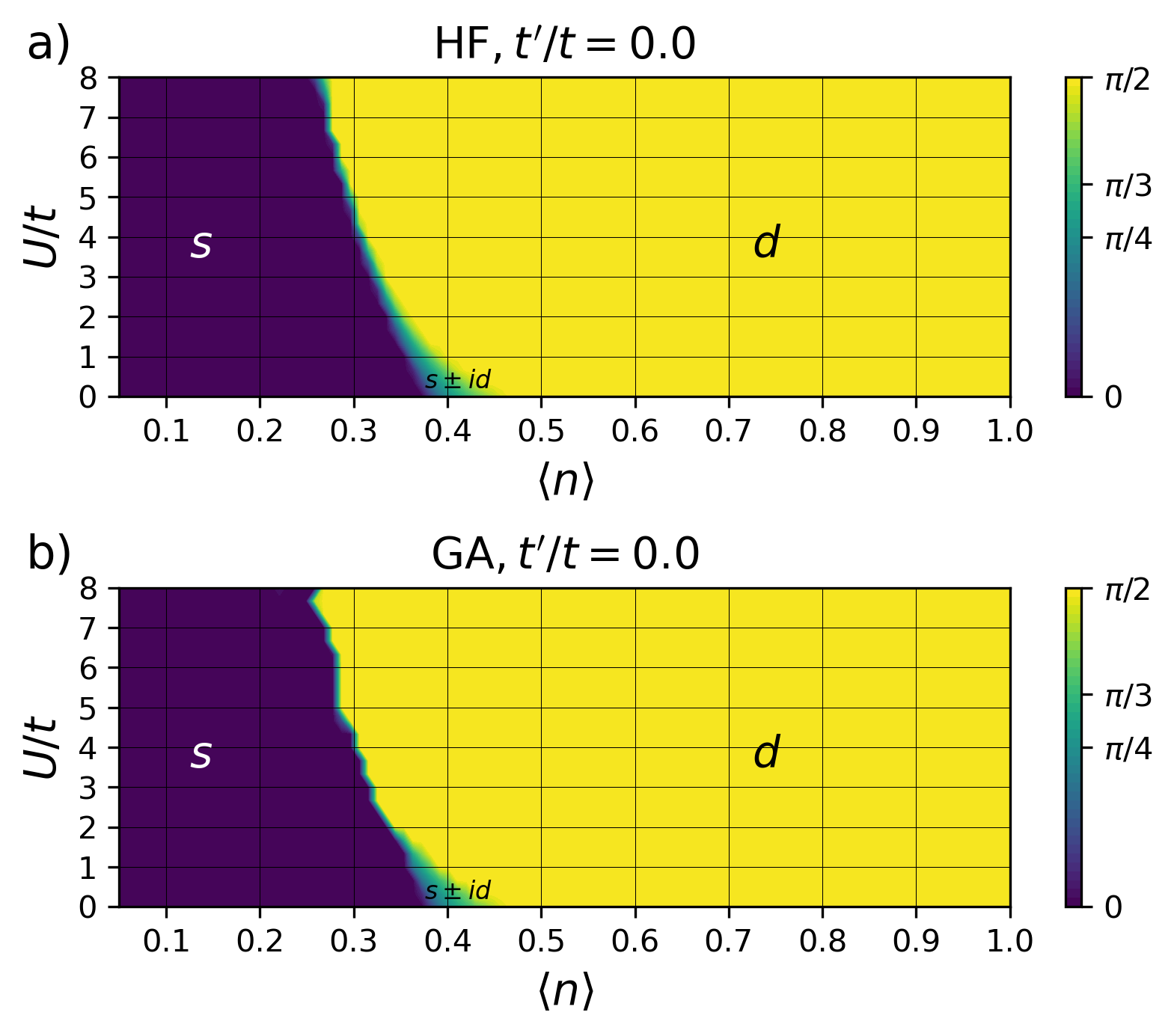}
     \caption{HF (a) and GA (b) phase diagrams for the phase
     $\alpha = \arctan(\Delta_d/\Delta_s)$ between s-wave (purple) and d-wave (yellow) states for next-nearest
     neighbor hopping $t'/t=0.0$}
     \label{fig1}
\end{figure}

\begin{figure}
     \centering
     \includegraphics[width=0.47\textwidth]{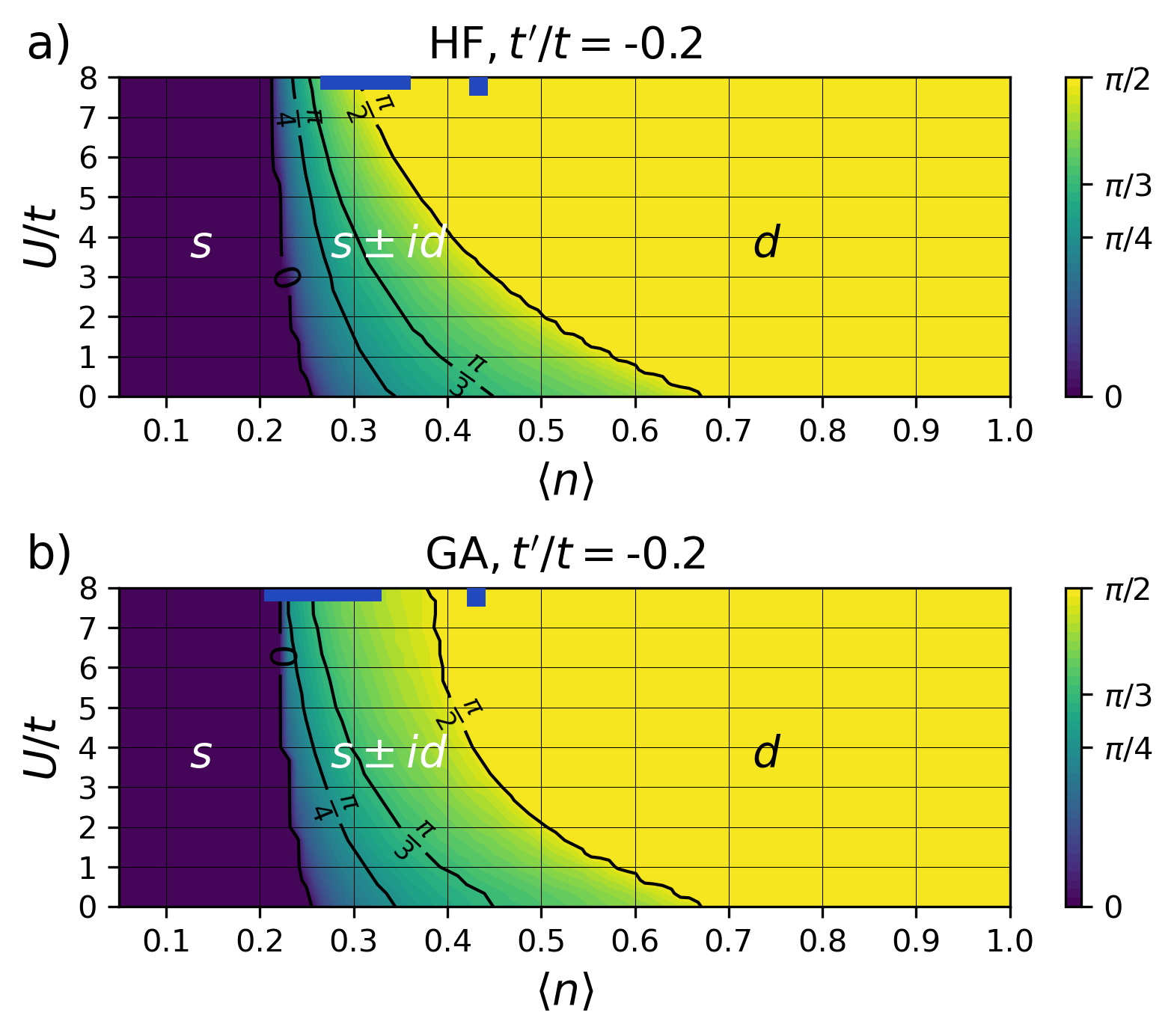}
     \caption{HF (a) and GA (b) phase diagrams for the phase
     $\alpha = \arctan(\Delta_d/\Delta_s)$ between s-wave (purple) and d-wave (yellow) states for next-nearest
     neighbor hopping $t'/t=-0.2$. In the multiple impurity case, cf. Sec. \ref{sec:multiple}, an average doping $\langle n\rangle= 0.44$ (blue square) leads to a charge distribution on the impurity sites ($V_i^{imp}/t=2$) which is indicated by the blue bar.}
     \label{fig3}
\end{figure}

We evaluate the lowest energy solution of the homogeneous system in $k$-space for the general $s \pm id$-state, where the complex order parameter has the structure $\Delta_{\pm x} \equiv \Delta_s +i\Delta_d$ and $\Delta_{\pm y} \equiv \Delta_s - i\Delta_d$, with real values $\Delta_s$, $\Delta_d$. From the corresponding self consistency equations, see appendix \ref{sec:GAFORM}, we determine the parameter $\alpha = \arctan(\Delta_d/\Delta_s)$  which allows to characterize the extended s-wave ($\alpha=0$), the
d-wave ($\alpha=\pi/2$) and $s+id$ regime ($0<\alpha<\pi/2$), yielding the phase diagrams shown in Figs. \ref{fig1}, \ref{fig3} for both, HF and GA approximation and two values of the next-nearest neighbor hopping $t'/t=0$ and $t'/t=-0.2$. Results for the case $t'/t = -0.4$ are discussed in appendix \ref{sec_tp04}. Calculations have been done for lattices up to $300\times 300$.

In general, the $s+id$ phase is located between the d-wave phase at large charge carrier concentration and the extended s-wave phase at low $\langle n \rangle$.
In fact, for $t'/t=0$ and $t'/t=-0.2$ the Fermi surface (FS) at $\langle n\rangle=1$ coincides with the nodes of the s-wave
structure factor $\gamma_k^s=2\left(\cos k_x +\cos k_y\right)$ so that the system cannot gain energy by opening a SC gap in the s-wave channel. In contrast, the structure factor for the d-wave state $\gamma_k^d=2\left(\cos k_x -\cos k_y\right)$ only intersects at four points with the FS so that in this case a substantial energy gain is possible via formation of the d-wave state. At low doping, where the FS is small and quasi circular one has still four intersections with the d-wave form factor (which is small in this region of k-space) but a significant distance to the nodes of $\gamma_k^s$, thus favoring the extended s-wave solution. At intermediate charge concentration $\langle n\rangle \sim 0.4$ s- and d-wave states become degenerate thus favoring the s+id solution in this region of the phase diagram. 
Upon starting in the $s+id$ phase and increasing $U$ 
leads to a transition towards the d-wave regime.
This is due to the local anomalous correlations $\langle c_{i\downarrow}c_{i\uparrow}\rangle$ which are zero for d-wave but finite for extended s-wave symmetry. In the latter case they get suppressed with $U$ thus removing the quasi-degeneracy in favor of d-wave symmetry. In contrast to HF, where the suppression of these correlations is the only mechanism to minimize the correlation energy, this is supported in the GA via the variational double occupancy parameter $\Tilde{D}$. Therefore, the $s+id$ state in GA extends to larger values of $U$ for a given doping than in HF. This is especially apparent for larger $|t'/t|$.

Increasing the magnitude of $t'/t$ induces a hole-like FS at large charge concentrations which no longer coincides with $\gamma_k^s$. Along with this, $k_F$ of the nodal FS points decreases and therefore also $\gamma_k^d$ (in magnitude). As a consequence, s-wave SC states will be stabilized whereas the condensation energy of d-wave states decreases, so that overall quasi-degeneracy (and thus the $s+id$ regime) extends towards larger $\langle n\rangle$. A similar argument can also be made for low charge concentrations. 
The resulting increase of the $s+id$ regime with increasing $|t'/t|$ is in agreement with the results of \cite{sid_phase_diagrams_paper} for $U/t=0$ case. It should also be noted that for $U/t=0$ and $t'/t=0$ the phase diagram coincides with the results from Refs.\cite{Clara_PhysRevB.105.014504,mueller_prb14}.

\subsection{Single impurity}\label{sec:single}
For a  homogeneous $s+id$ state the phase of the complex
order parameter $|\Delta|e^{i\varphi}$ can be
gauged away by the transformation $c_{i\sigma}\to c_{i\sigma}e^{-i\sigma\varphi(i_x-i_y)}$. In the kinetic
energy this transformation acts as a spin dependent vector
potential $\vec{A}_\sigma \sim \sigma\varphi(1,-1)$ signaling the emergence of TRSB. Thus, the induced currents in the spin up and spin down channel exactly compensate each other.

\begin{figure}
    \centering
    \includegraphics[width=0.85\linewidth]{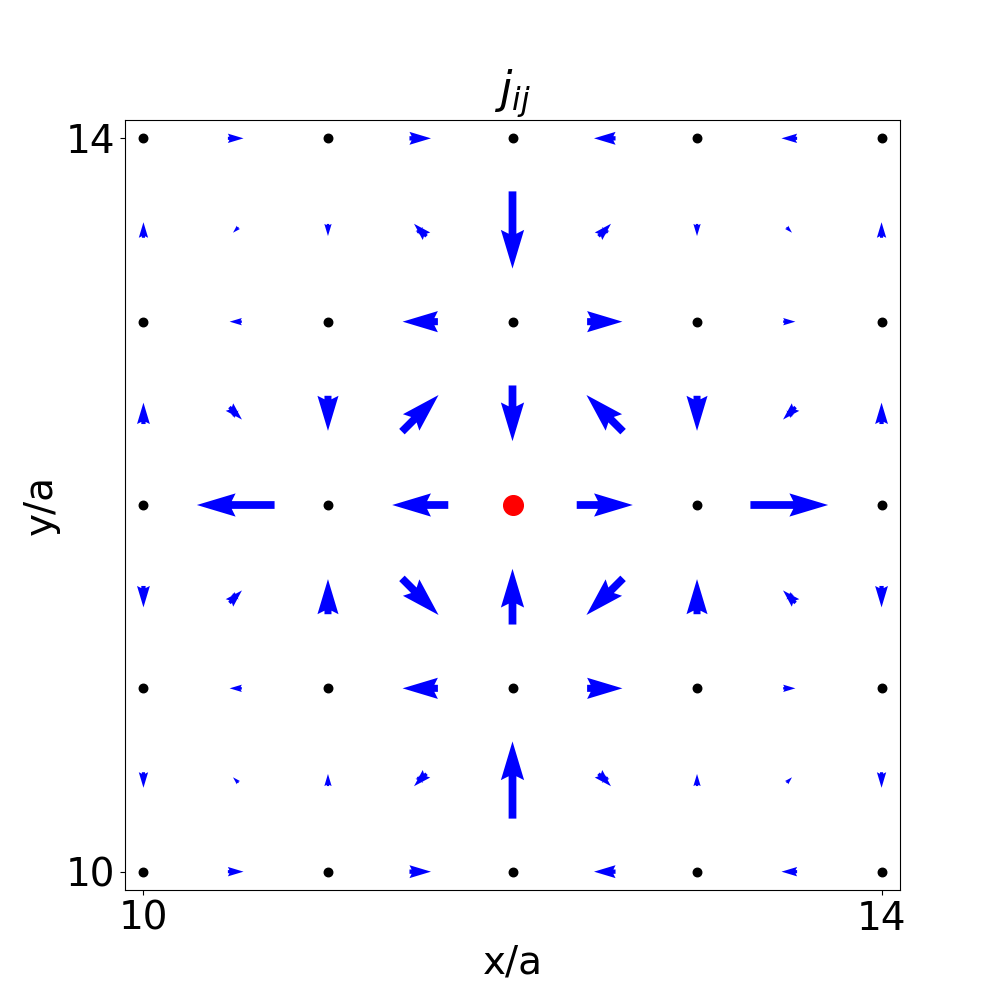}
    \caption{GA current pattern in an area around a single impurity within a $25 \times 25$ lattice. Parameters: $U/t = 6$, $t'/t=-0.2$, $\langle n\rangle=0.35$. Impurity  strength $V_{\text{imp}}/t=2$. }
    \label{fig5}
\end{figure}

This is no longer the case when one considers a defect in a chiral superconductor which induces the formation of loop currents \cite{PhysRevB.94.064519, PhysRevLett.112.017003, PhysRevB.101.054507, PhysRevLett.102.217002, PhysRevB.91.161102, PhysRevLett.116.097002, PhysRevB.104.094505, PhysRevB.106.214530, Clara_PhysRevB.105.014504}.
In particular, for a $s \pm id $-state,
which breaks $C_4$ symmetry but is invariant under the combination of $C_4$ rotation and time reversal, the current pattern takes the form shown in Fig. \ref{fig5}, where the current flow towards the impurity along the $y$-direction is compensated by currents away from the impurity along $x$.
Symmetry also dictates cancellation of the net vorticity and the vanishing of the current flow
in diagonal direction outwards from or inwards to the impurity.
We have checked that in our computations current conservation is obeyed at each node.

For a single impurity the emerging currents $j_{ij}$ occur mainly in the $s+id$ regions of the homogeneous system, cf. Figs. \ref{fig1},\ref{fig3},\ref{fig4}.
However, due to the enhanced (reduced) density in the vicinity of the impurity for negative (positive) $V_i^{imp}$ currents are
also induced slightly below (above) the $s+id$ regime of the homogeneous system.

\begin{figure}
     \centering
     \includegraphics[width=0.47\textwidth]{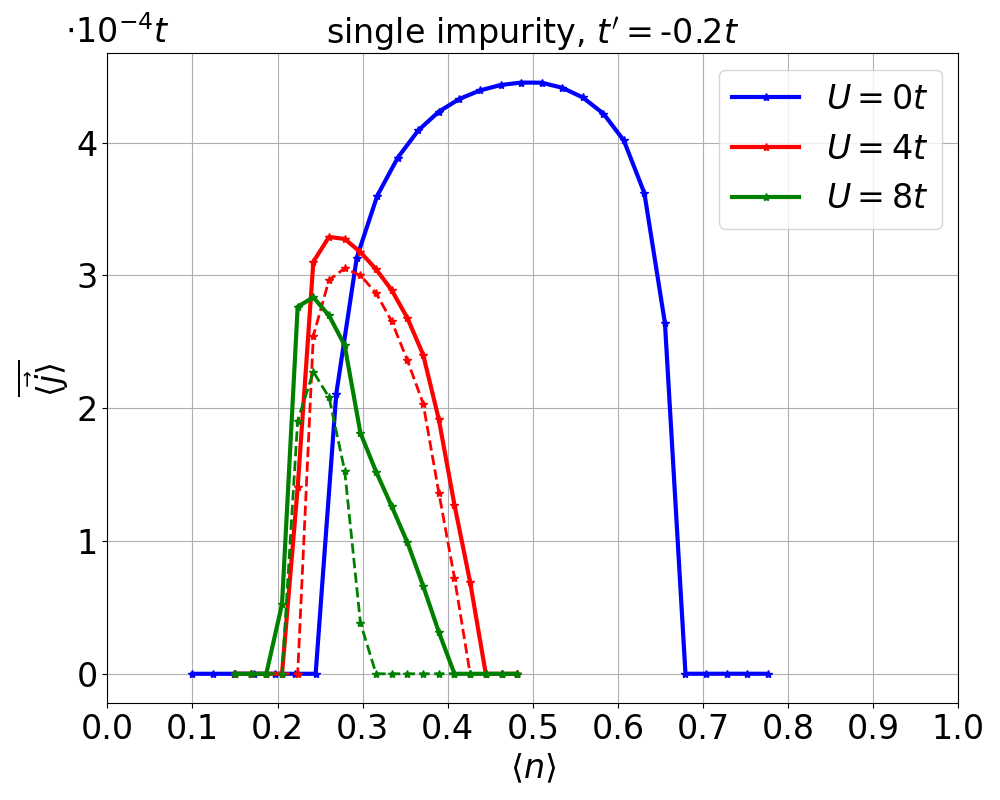}
     \caption{Averaged current strengths for $t'/t=-0.2$ in presence of one single impurity. Dashed lines are the HF-results, full lines are obtained from GA. Currents are obtained only in the $s\pm id$-regions, which have been determined in the phase diagram figure \ref{fig3}. The gridsize here is 40x40 and $V_{\text{imp}}/t = 2$.}
     \label{fig7}
\end{figure}

Further on, we define an average current via
$\langle j \rangle \equiv \frac{1}{N}\sum_{ij} |j_{ij}|$ where $N$ is the number of lattice sites and $ij$ denotes each pair of sites connected by $t$ and $t'$, respectively.
Figure \ref{fig7} shows the dependence of 
$\langle j \rangle$ on the carrier concentration $\langle n\rangle$ for $t'/t=-0.2$ and various $U/t$.
Within a Ginzburg-Landau analysis \cite{PhysRevLett.102.217002,PhysRevB.91.161102,breio_phd} it can be shown that the total current depends on both, phase and amplitude fluctuations around the impurity. Therefore the maximum current in Fig. \ref{fig7} does not
coincide with the density of the $\alpha=\pi/4$ regime in
Fig. \ref{fig3} but is also influenced by the
spatial variation of the order parameter around the impurity.

Increasing $U/t$ shrinks the $s+id$ region
towards lower densities, cf. Fig. \ref{fig3}, with concomitant smaller amplitude variations of the SC order parameter and thus a reduced maximum total current. 
For the considered densities the impact of correlations beyond HF is small so that GA and HF yield similar currents. 
It should also be noted that removing the diagonal currents from the total current only leads to an overall reduction of the curves
in Fig. \ref{fig7}, but does not alter significantly the shape of the curves or the ratio between them. 



\subsection{Multiple impurities}\label{sec:multiple}

\begin{figure}
     \centering
     \includegraphics[width=0.47\textwidth]{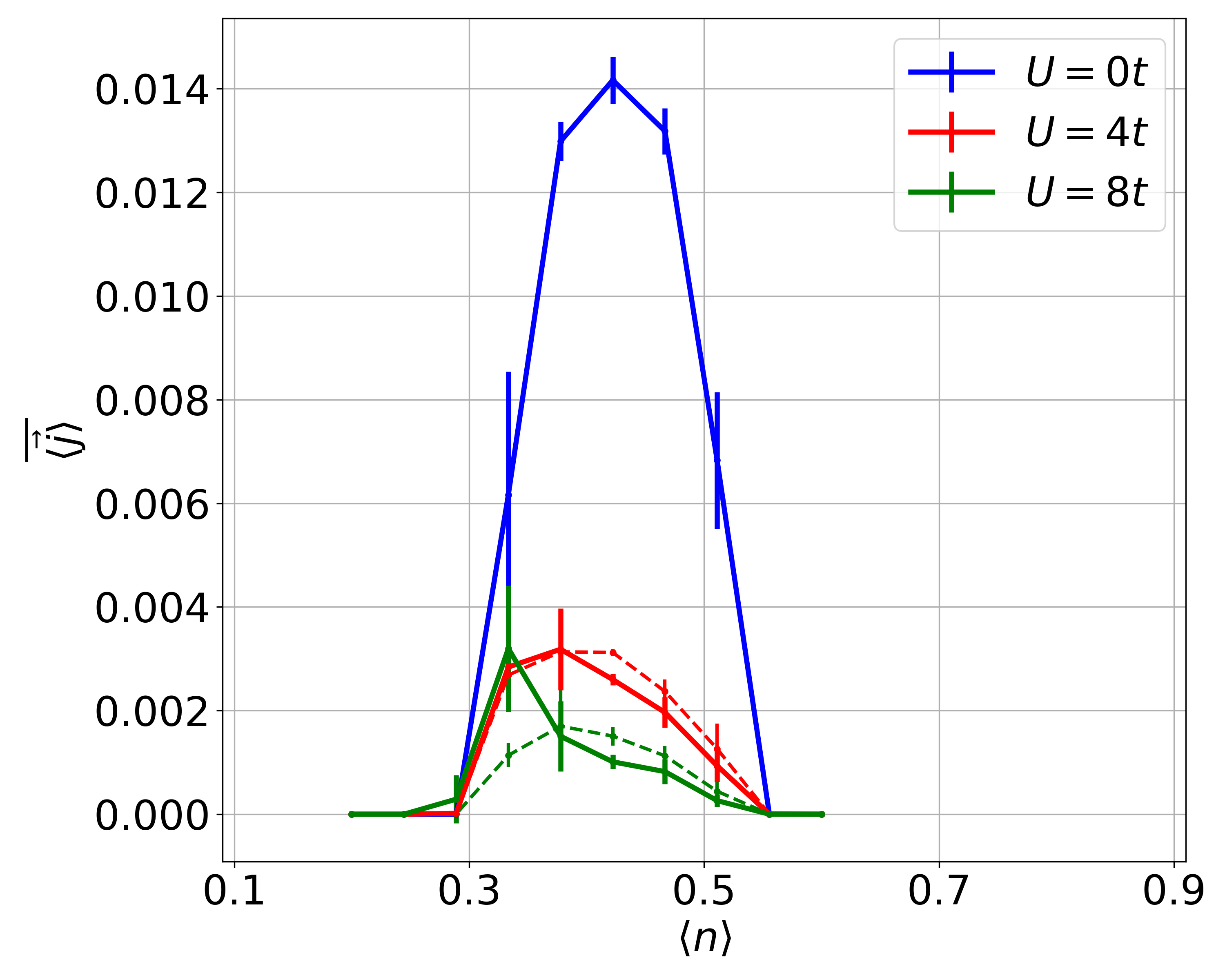}
     \caption{Current strengths for $t'/t=0.0$ in presence of multiple impurities. Dashed lines are the HF-results, full lines are obtained from GA.}
     \label{fig10}
\end{figure}

\begin{figure}
     \centering
     \includegraphics[width=0.47\textwidth]{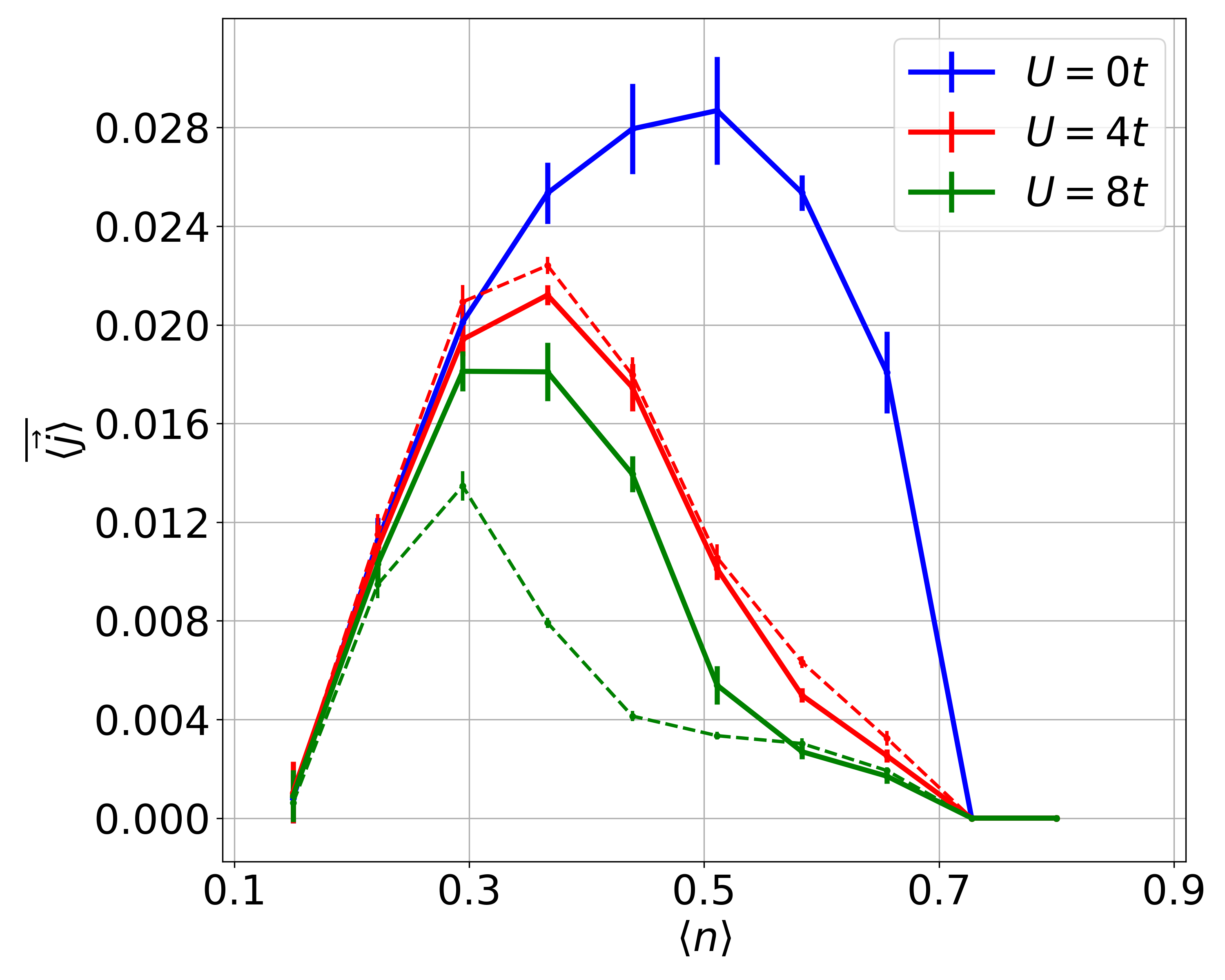}
     \caption{Current strengths for $t'/t=-0.2$ in presence of multiple impurities. Dashed lines are the HF-results, full lines are obtained from GA.}
     \label{fig11}
\end{figure}

We proceed by analyzing the case of a finite impurity density, where we adopt the scenario of Refs. 
\cite{Kivelson,Clara_PhysRevB.105.014504}, i.e. each impurity corresponds to doping of one hole. Thus, the impurity concentration $c$ is related to the charge carrier concentration $n$ via $c = 1-n$. Impurities are placed randomly on sites of a $35\times 35$ grid and in the following we
take $V^{\text{imp}}_i = 2t$ for the impurity strength.
As in the previous section, we calculate the mean currents $\langle j \rangle$ for values $n$, $U$ and $t'$, now for $5$ randomly created impurity configurations, of which the average and standard deviation is determined. The corresponding results are shown in Figs. \ref{fig10}, \ref{fig11}. 

Due to the inhomogeneous charge distribution, induced by
the impurities, local $s\pm id$-regions, associated with a finite current, are now obtained
beyond the corresponding regime of the homogeneous system similar to Refs. \cite{Kivelson,Clara_PhysRevB.105.014504}.
In contrast to the single-impurity case, Fig. \ref{fig7}, where the $\langle j \rangle\ne 0$ regime is dependent on $U/t$ we find, that for a finite impurity concentrations the corresponding doping range becomes independent of the local repulsion. The total current strength becomes suppressed 
upon increasing $U/t$ for both, HF and GA.
However, in particular for finite $t'/t$ and large $U/t$ we observe a strong enhancement of the currents obtained from the Gutzwiller approach as compared to the corresponding HF result (cf. green curves in Fig. \ref{fig11}).

\begin{figure}
     \includegraphics[width=0.5\textwidth]{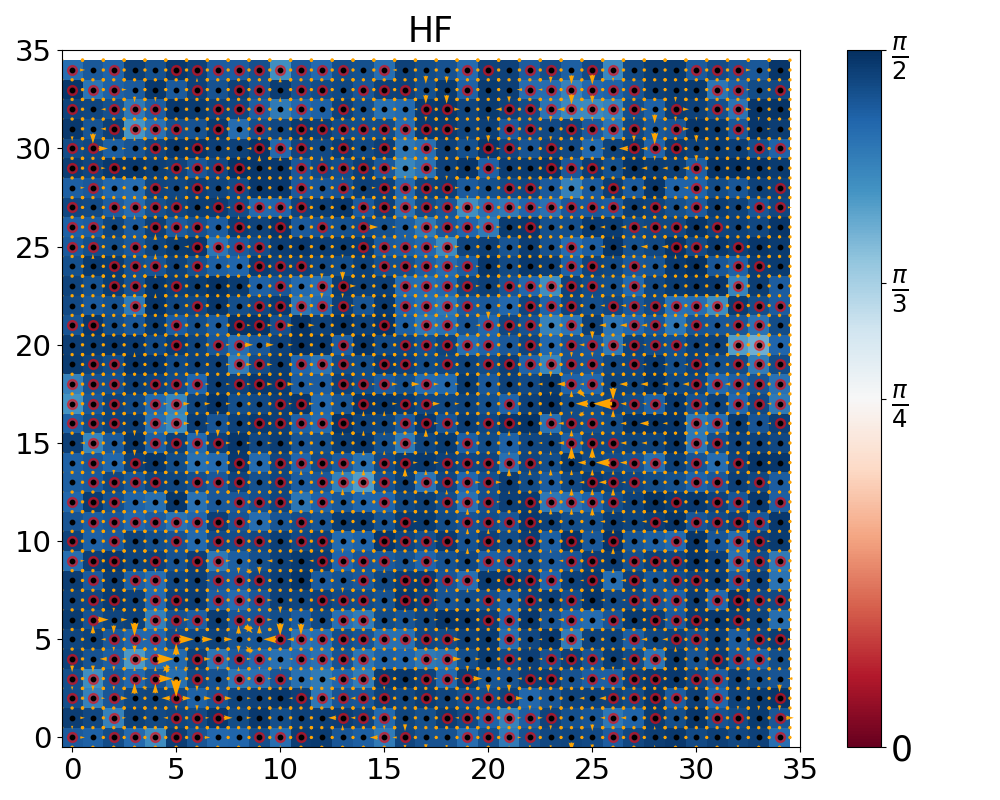}
     \includegraphics[width=0.5\textwidth]{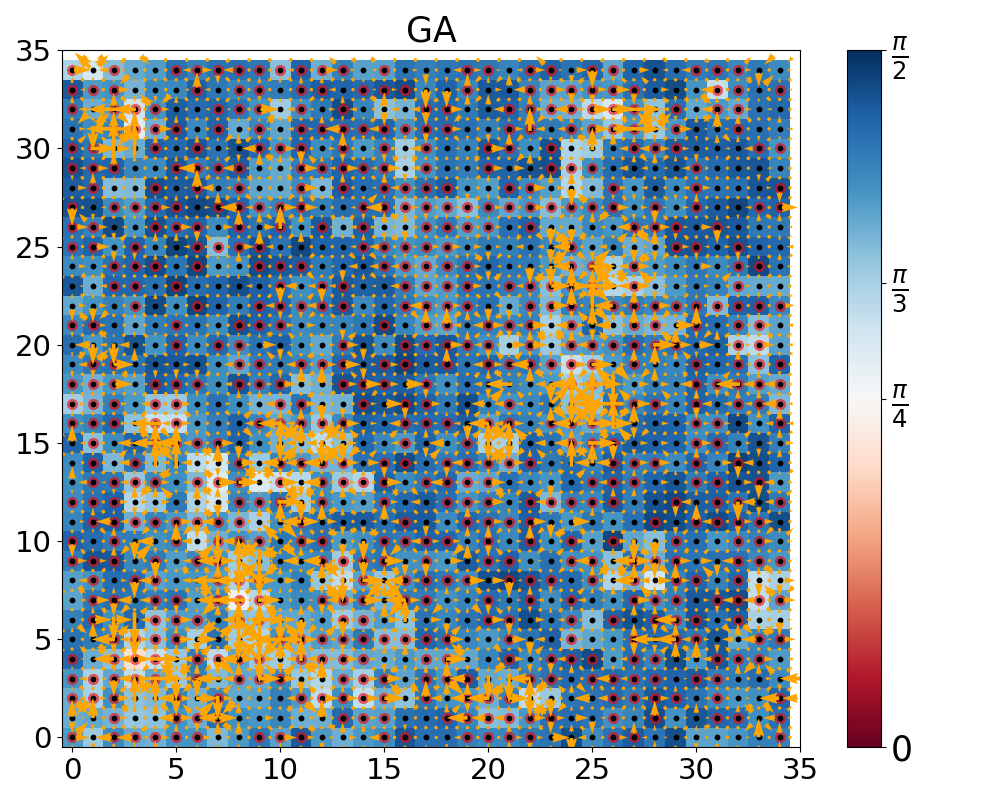}
     \caption{Local phase between d- and extended s-wave OP together with the induced currents obtained within HF (upper panel) and GA (lower panel). Current arrows are only shown above a minimum value. We have checked that current conservation is obeyed at each node. Parameters: charge concentration $\langle n\rangle=0.44$, $U/t=8$, $t'/t=-0.2$.}
     \label{fig12}
\end{figure}

Consider e.g. a charge concentration $\langle n\rangle= 0.44$ (indicated by blue squares in Fig. \ref{fig3}) for which at $t'/t=-0.2$ and $U/t=8$ the total current evaluated within GA exceeds by a factor of $\sim 3$ the corresponding HF result. 
For both HF and GA, Fig. \ref{fig12} reports the local phase between extended s-wave and d-wave order parameters $\alpha_n = \text{arctan}(|\Delta_d(n)/\Delta_s(n)|)$. Clearly, currents appear predominantly in local $s+id$ regions (with $\alpha_n \approx \pi/4$) which manifest at the impurity sites. We find that in the GA these sites are also characterized by a sizeable value of the local order parameter $\Delta_n$ which in the HF solution is significantly smaller on those sites, cf. figure \ref{fig15} in appendix B. This is in agreement with our argument from Sec. \ref{sec:hom} that the GA can sustain the $s+id$ state to larger values of $U/t$ whereas in HF the more rapid suppression of $\Delta_n$ with $U/t)$ induces the transition towards the pure d-wave state (blue regions in Fig. \ref{fig12}). The configuration-averaged Pearson correlation coefficients between the currents $\vec{j}_i$ and the quantities $\sin\left(2 \alpha_n \right)$ (where the maximum value is reached for $\alpha_n = \frac{\pi}{4}$) or respectively Re$\Delta_i$ for this case (i.e. GA, $U/t=8$, $\langle n\rangle= 0.44$, $t'/t = - 0.2$) are $\text{corr}\left(\vec{j}, \sin\left(2\alpha_n\right)\right) = 0.36 \pm 0.05$ and $\text{corr}(\vec{j}_i,\text{Re}\Delta_i) = 0.51 \pm 0.08$, which strongly supports that hypothesis. Consistently, the variation of the charge density on the $s+id$ impurity sites, indicated by a blue bar in Fig. \ref{fig3} is shifted towards the d-wave region in HF as compared to GA where it is almost completely within the $s+id$ phase.

The similar argument also accounts for the doping dependent current strength in case of $t'/t=0$, see Fig. \ref{fig10}.
In the homogeneous case and upon increasing $U/t$ the $s+id$ phase is confined to a narrow area between d- and s-wave regime. In the inhomogeneous system the concomitant charge
carrier distribution causes the density on some of the impurity sites to match with the density where TRSB is favored.
With regard to the homogeneous system this enlarges the $s+id$
regime to concentrations $0.3 \lesssim \langle n \rangle \lesssim 0.55$ for the considered values of $U/t$, however, as the TRSB density window is much smaller than in case
of $t'/t=-0.2$, the resulting current strength is correspondingly reduced.

\section{Summary and Conclusions}\label{sec:sum}

In this work, we investigated the influence of local correlations on the loop currents emerging from impurities in a one-band Hubbard model supplemented with a nearest-neighbor
attractive interaction as relevant for cuprate superconductors. As has already been observed in earlier studies for the case $U/t=0$ \cite{Kivelson,Clara_PhysRevB.105.014504}, disorder promotes loop currents in a doping range that extends far beyond the $s+id$ region of the homogeneous system. This result persists in the presence of local correlations, however, with loop currents which are reduced upon increasing $U$. Interestingly, the incorporation of correlations beyond HF within the Gutzwiller approximation makes this suppression less pronounced since it supports the persistence
of local pairing correlations, and thus the possibility of an $s+id$ phase, up to larger values of $U$. We have also shown that the loop current phase becomes even more extended upon considering next-nearest neighbor hopping in the hamiltonian.

It is interesting to estimate the magnetic moment $m = I\cdot F$ associated with the
loop currents. From our investigations such currents should be in principle observable
in cuprate materials with large $|t'/t|\approx 0.4$, cf. Fig. \ref{fig13}, as
e.g. HgBa$_2$Ca$_2$Cu$_3$O$_8$ or Tl$_2$Ba$_2$Ca$_2$Cu$_3$O$_{10}$ \cite{PhysRevLett.87.047003}.
The nearest-neighbor hopping is taken as $t=350 meV$ and we evaluate the largest current from the (triangle) loop at the impurity, cf. figure \ref{fig5}, with area $F=\frac{a^2}{2}$.
For $U/t=0$ one obtains a magnetic moment of $\sim {\cal O}(10^{-2})$ Bohr magnetons 
which for $U/t=6$ is then reduced to $\sim {\cal O}(10^{-3})$ Bohr magnetons.
This is by one (two) magnitudes smaller than the magnetic moment observed by polarized
neutron scattering \cite{PhysRevLett.96.197001}, and which has been attributed to the loop current order proposed in Ref. \onlinecite{varma97}. Furthermore, the experimentally observed moment shows a tilt of $\sim 45^\circ$ with respect to the c-axis whereas the moments from the loop currents in the present model point exactly along the c-direction.

It would therefore be interesting to investigate a similar scenario within the three-band model and to see whether the induced loop currents, at least locally around the impurities, resemble the configuration proposed in Ref. \onlinecite{varma97}. Moreover, also a possible
loop current flow involving apical oxygens could be included in such a consideration in order
to resolve the discrepancy with regard to the orientation of the moments.
Work in this direction is in progress.

\section*{Acknowledgements}

We thank Chiranjit Mahato for stimulating discussions.

\appendix

\section{The case $t'/t=-0.4$}\label{sec_tp04}
\begin{figure}
     \centering
     \includegraphics[width=0.47\textwidth]{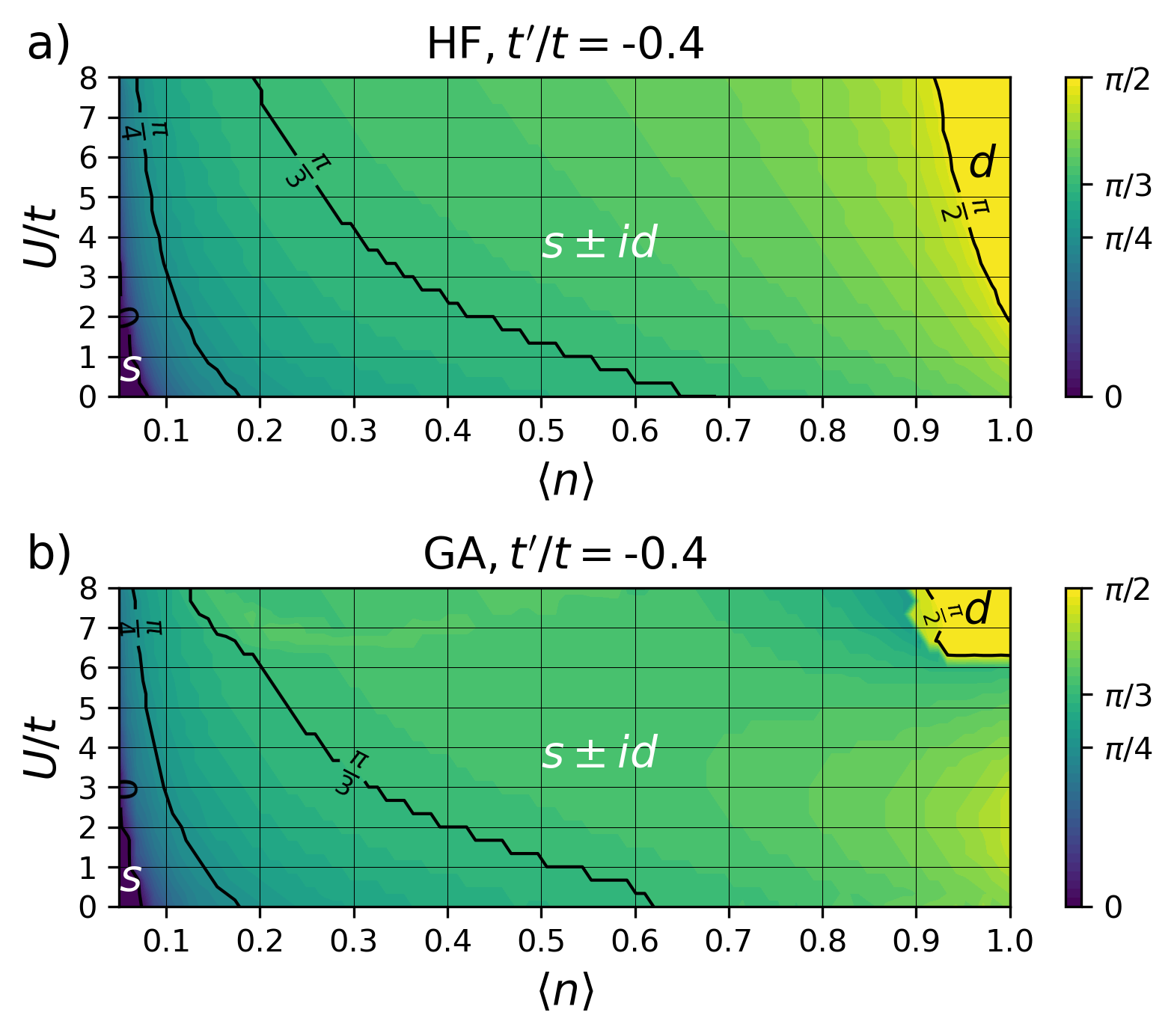}
     \caption{HF (a) and GA (b) phase diagrams for the phase
     $\alpha = \arctan(\Delta_d/\Delta_s)$ between s-wave (purple) and d-wave (yellow) states for next-nearest
     neighbor hopping $t'/t=-0.4$}.
     \label{fig4}
\end{figure}

In case of a nearest-neighbor hopping $t'/t=-0.4$ the $s+id$ covers most of the phase diagram
as can be seen from Fig. \ref{fig4}. Only for charge concentrations $\langle n\rangle \lesssim 0.1$ the extended s-wave solution is recovered whereas the pure d-wave is stable around
half-filling. In particular, within GA the latter only is found above a critical value
of $U/t\approx 6$ whereas in HF it extends to far smaller values of the onsite repulsion. Furthermore, in HF the phase $\alpha = \arctan(\Delta_d/\Delta_s)$ stays close to the value of $\alpha=\pi/2$
over a significant doping range away from half-filling.

\begin{figure}
    \centering
    \includegraphics[width=0.85\linewidth]{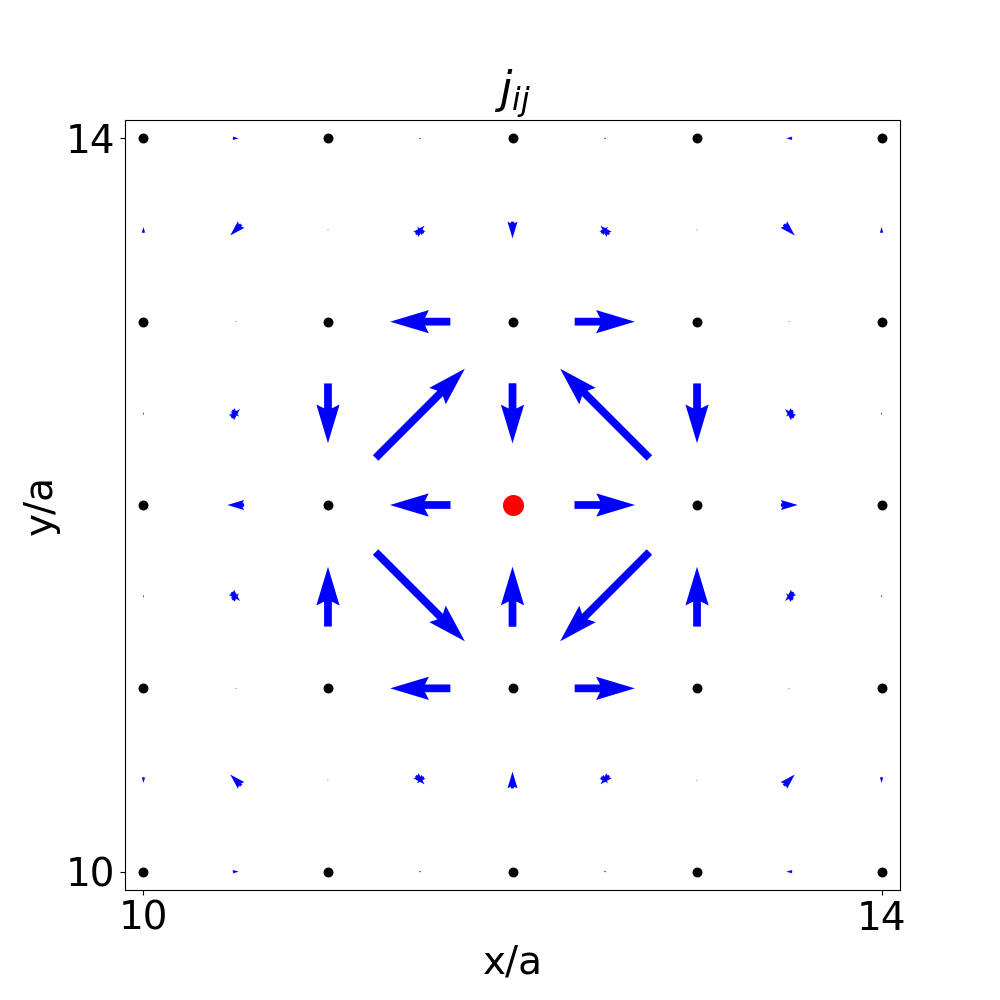}
    \caption{GA current pattern in an area around a single impurity within a $25 \times 25$ lattice. Parameters: $U/t = 6$, $t'/t=-0.4$, $n=0.35$. Impurity  strength $V_{\text{imp}}/t=2$. Compared to the single-impurity case with $t'/t=-0.2$, cf. figure \ref{fig5}, here the diagonal currents dominate.}
    \label{fig_stroeme}
\end{figure}

\begin{figure}
     \centering
     \includegraphics[width=0.47\textwidth]{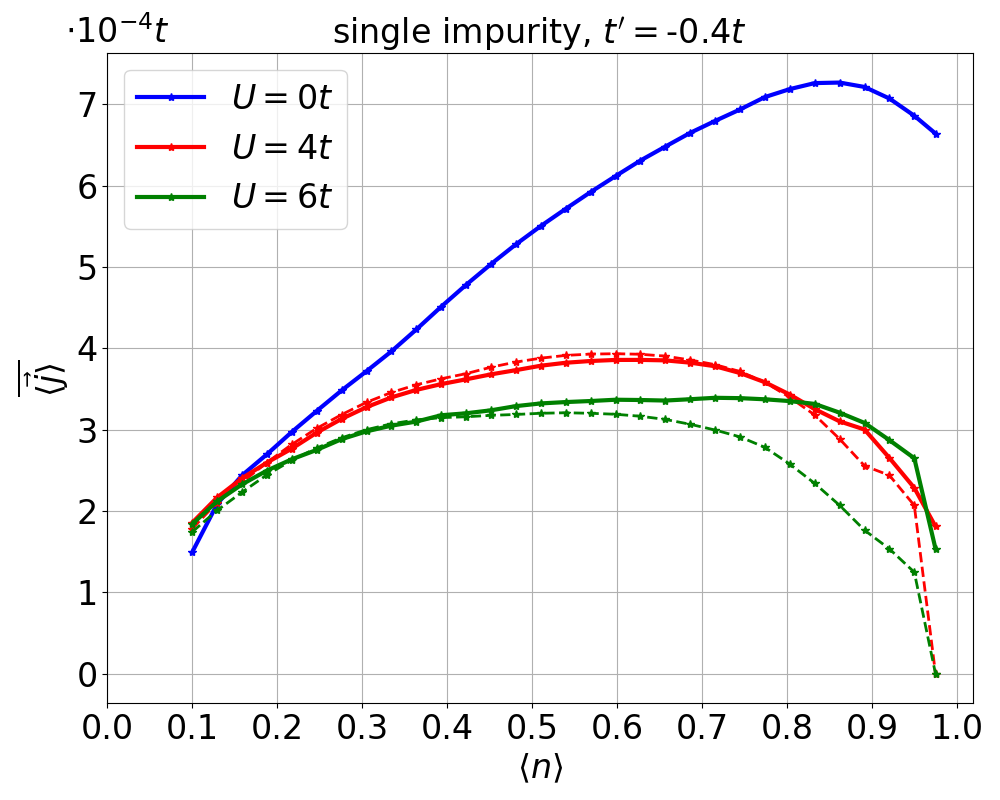}
     \caption{Averaged current strengths for $t'/t=-0.4$ in presence of one single impurity as function of charge concentration. Dashed lines are the HF-results, full lines are obtained from GA. Currents are obtained only in the $s\pm id$-regions, which have been determined in the phase diagram figure \ref{fig3}. The gridsize here is 40x40 and $V_{\text{imp}}/t = 2$.}
     \label{fig13}
\end{figure}

\begin{figure}
     \centering
     \includegraphics[width=0.47\textwidth]{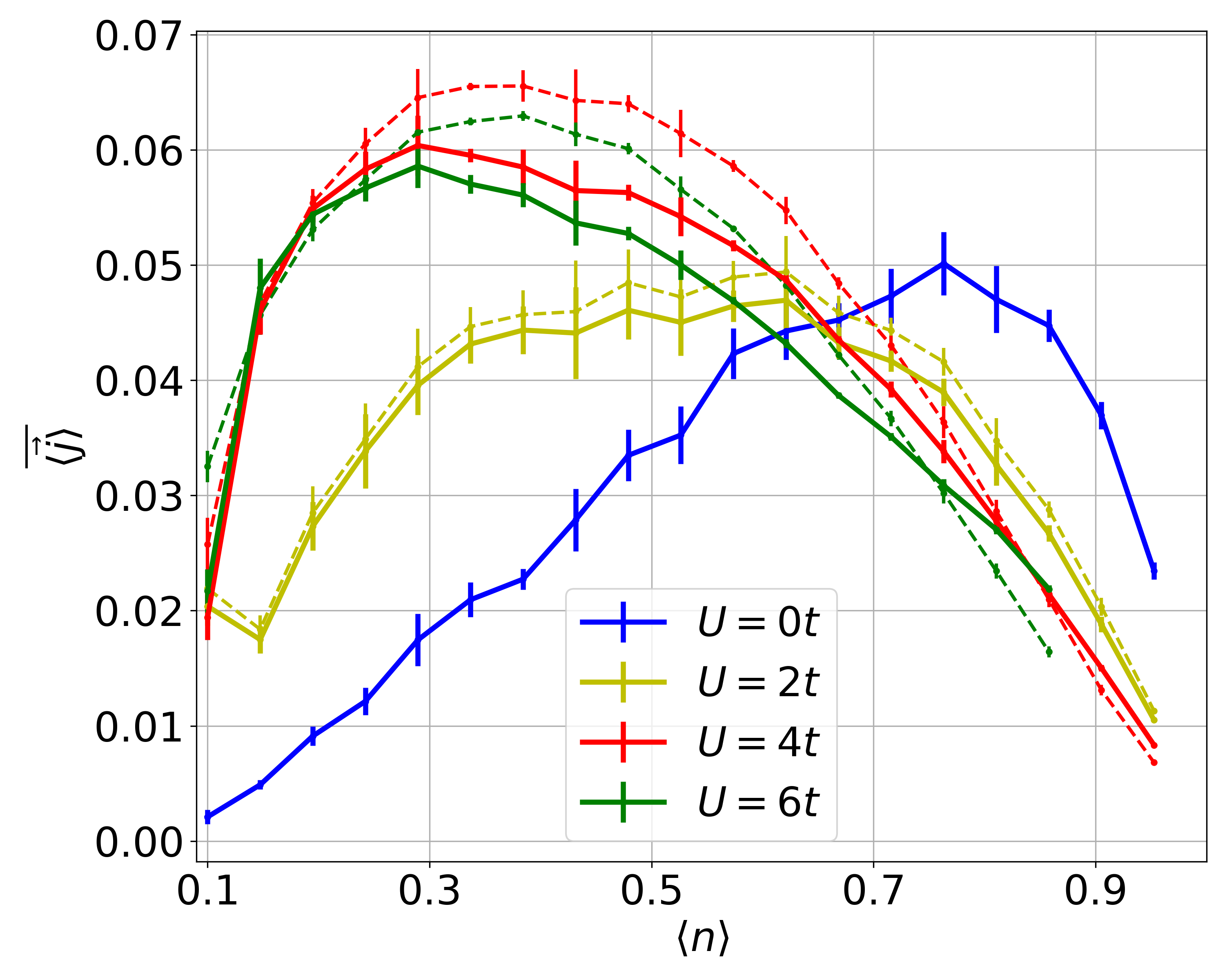}
     \caption{Current strengths for $t'/t=-0.4$ in presence of multiple impurities with $V_{\text{imp}}/t = 2$ as a function of charge concentration. Dashed lines are the HF-results, full lines are obtained from GA.}
     \label{fig14}
\end{figure}

This has profound consequences for the average current strength in the single impurity case
as shown in Fig. \ref{fig13}. In particular, for large $U/t=6$ the HF currents decrease much more rapidly upon approaching $\langle n\rangle=1$
than in case of the GA. The same feature, although less pronounced, is observed in the multi-impurity case, Fig. \ref{fig14}. It should be noted
that for $t'/t=-0.4$ and doping close to half-filling the major difference
between HF and GA arises from the currents 
between next-nearest neighbors adjacent to the impurity, cf. figure \ref{fig_stroeme}. 

In fact, consider a concentration $\langle n\rangle=0.86$, then the charges are pushed away from the impurities ($V_{imp}/t=2$), so that adjacent sites acquire an average density $\langle n\rangle=0.9$. Thus, similar to our argument in 
case of $t'/t=-0.2$ for large $U/t$ the corresponding point in the HF phase diagram Fig. \ref{fig4} is closer to the d-wave phase than in GA so that in the former the corresponding currents are reduced.

\section{What drives the current formation?}
\begin{figure}
     \includegraphics[width=0.5\textwidth]{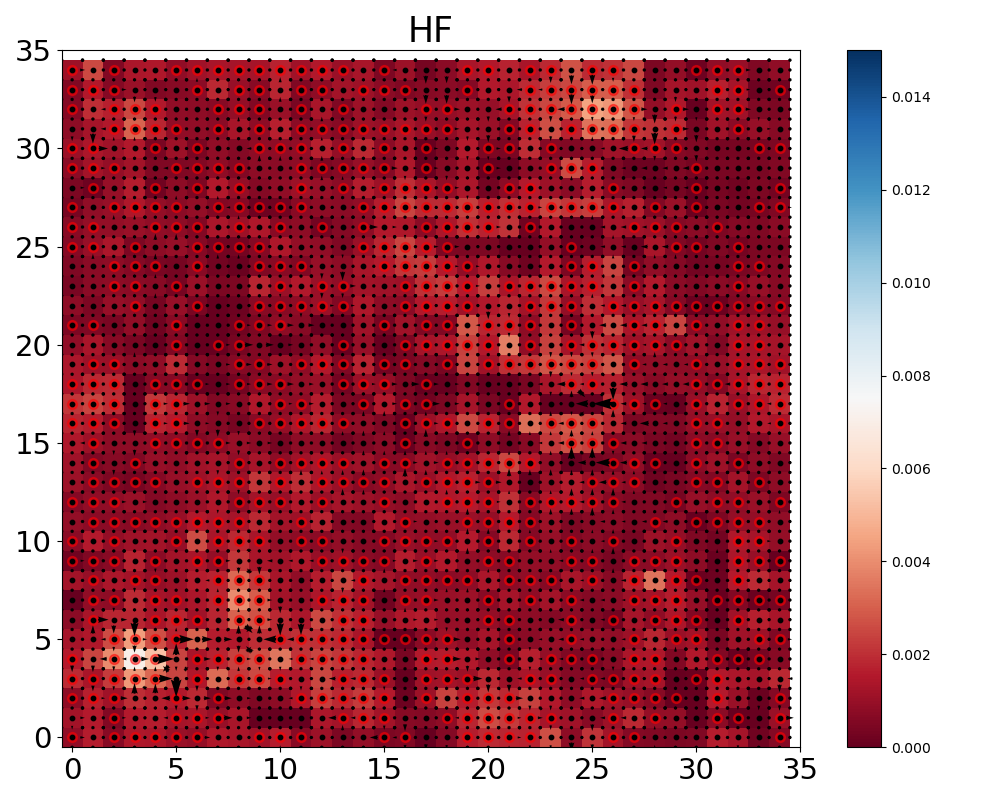}
     \includegraphics[width=0.5\textwidth]{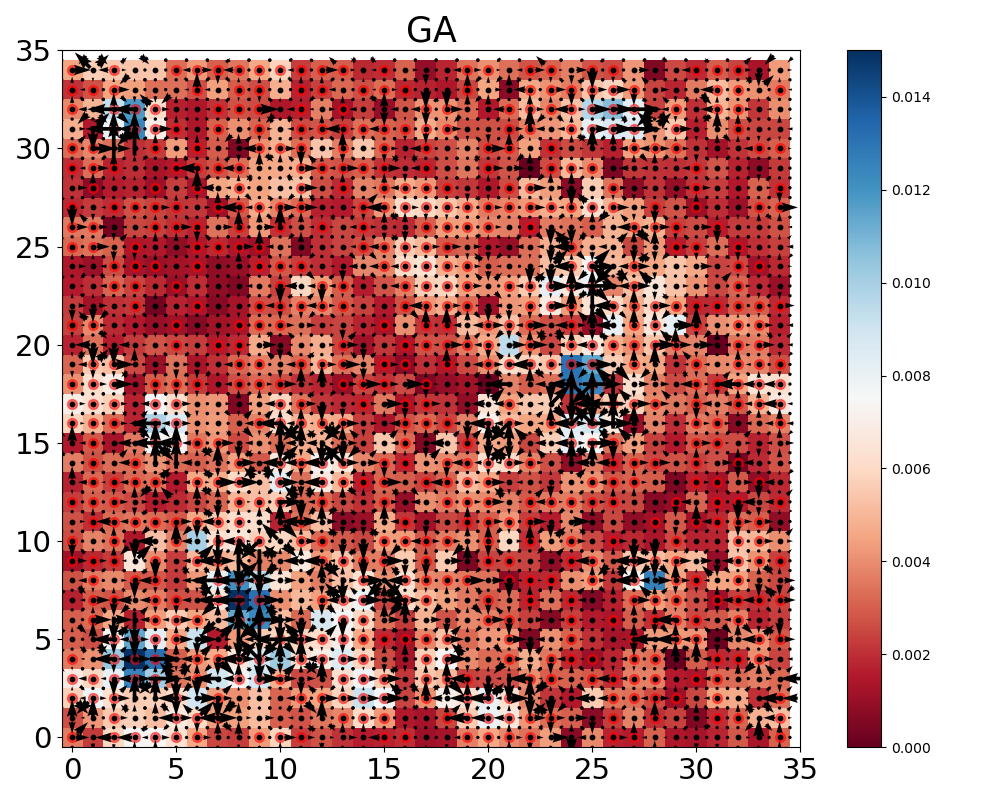}
     \caption{Real part of the local order parameter $\Delta_n$ together with the induced currents obtained within HF (upper panel) and GA (lower panel).
     Current arrows are only shown above a minimum value. We have checked that current conservation is obeyed at each node. Parameters: charge concentration $\langle n\rangle=0.44$, $U/t=8$, $t'/t=-0.2$.}
     \label{fig15}
\end{figure}

In Sec. \ref{sec:multiple} we have shown that in the disordered system currents
are induced in regions with a locally sizeable $s+id$ order parameter.
We have further argued that GA can sustain these regions to larger values
of $U/t$ due to the fact, that the minimization of the double occupancy is less dependent of a suppression of local anomalous correlations than in HF where it is responsible for the suppression of the s-wave contribution upon
increasing $U/t$. In order to substantiate this hypothesis we show in Fig. \ref{fig15} the real part of the local correlations together with the emerging currents (note that $\text{Im} (\Delta_n)$ does not correlate with the currents).

For the chosen parameters ($U/t=8$, $\langle n\rangle =0.44$, $t'/t=-0.2$) the GA
result displays numerous regions (in blue) with sizeable $\text{Re}(\Delta_n)$ which also
host the dominant currents. In contrast, the number of such regions is rare in HF (although the disorder configuration is the same) and thus the concomitant current formation is suppressed.

We finally note that it has been shown that for $U/t=0$ the quantity $|\sin(\phi^d_i - \phi^s_i)|\cdot \frac{1}{4}\sum_{\delta} |\Delta_{i,i+\delta}|$ (where $\phi^{s,d}_i$ are the phases of the complex s-/d-wave order parameters $\Delta_{s,i}$, $\Delta_{d,i}$) correlates most with the current formation (see \cite{Clara_PhysRevB.105.014504}). Evaluation the corresponding configuration-averaged Pearson coefficient for the parameters of Fig. \ref{fig12}, \ref{fig15} one finds $\text{corr}(\vec{j}_i,|\sin(\phi^d_i - \phi^s_i)|\cdot \frac{1}{4}\sum_{\delta} |\Delta_{i,i+\delta}|) = 0.04 \pm 0.05$, i.e. a value
significantly smaller as compared to the correlation with the local
order parameter $\text{corr}(\vec{j}_i,\text{Re}\Delta_i) = 0.51 \pm 0.08$
as discussed in Sec. \ref{sec:multiple}.

\section{Gutzwiller formalism}\label{sec:GAFORM}

From \cite{GS_FB_08} we adopt the charge rotational invariant Gutzwiller approximation (GA), which is obtained by first rotating the operators

\begin{align*}
    J^x_i &= \frac{1}{2} \left( c^{\dagger}_{i\uparrow}c^{\dagger}_{i\downarrow} + c_{i\downarrow}c_{i\uparrow}  \right) \;, \\
    J^y_i &= -\frac{i}{2} \left( c^{\dagger}_{i\uparrow}c^{\dagger}_{i\downarrow} - c_{i\downarrow}c_{i\uparrow}  \right) \;, \\
    J^z_i &= \frac{1}{2} \left( c^{\dagger}_{i\uparrow} c_{i\uparrow} + c^{\dagger}_{i\downarrow} c_{i\downarrow} - 1  \right) \\
\end{align*}

into a system without superconducting order, i.e. $(0,0, \Tilde{J}^{z}_i)$. Then the Kotliar-Ruckenstein slave-boson-method is applied, with a mean-field approximation for the bosons. After this, it is rotated back to the original frame. All this leads to a renormalization of the creation- and annihilation operators $c_{i,\uparrow} = z_i f_{i,\uparrow}$ and $c^{\dagger}_{i,\downarrow} = z_i f^{\dagger}_{i,\downarrow}$ with

\begin{align*}
    z_{i} = \frac{\sqrt{E_i P_i} + \sqrt{P_i D_i}}{\sqrt{E_i + P_i} \sqrt{D_i + P_i }} \;.
\end{align*}

The slave-boson constraints are

\begin{align*}
    D_i + 2 P_i + E_i = 1 \;, \\
    P_i + D_i = \frac{1}{2} + \text{sign}(\langle J^z_i \rangle)\langle J_i \rangle \;,
\end{align*}

with 
\begin{eqnarray*}
\langle J_i \rangle &=& \sqrt{\langle J^x_i \rangle^2 + \langle J^y_i \rangle^2+ \langle J^z_i \rangle^2} \\
&=&\sqrt{|\Delta_i|^2 + \frac{1}{4}\left(n_{i,\sigma} - 1\right)^2}
\end{eqnarray*}
and
$\Delta_i=\langle J_i^x\rangle-i \langle J_i^y\rangle$, $n_{i,\sigma}=1+2\langle J_i^z\rangle$.

The mean double occupancy variable in this formalism can be derived as

\begin{align*}
    \Tilde{D}_i = D_i + \langle J^z_i \rangle - sign(\langle J^z_i \rangle)  \langle J_i \rangle \;,
\end{align*}
and to eliminate $E_i, P_i$ variables via the constraint so that the Gutzwiller
renormalization factors are given by

\begin{align*}
    z_i(\Tilde{D}_i) =& \frac{\sqrt{\frac{1}{2} - \Tilde{D}_i + \langle J^z_i \rangle} }{\sqrt{\frac{1}{4} - \langle J_i \rangle^{2} }} \left( \sqrt{\Tilde{D}_i -  \langle J^z_i \rangle - \langle J_i \rangle} \right. \\ 
    & \left. + \sqrt{\Tilde{D}_i -  \langle J^z_i \rangle + \langle J_i \rangle} \right) \;.
\end{align*}

Introducing Lagrange-multipliers $\lambda_{1,i}$, $\lambda_{2,i}$ for the local order parameter and the charge density leads to the following hamiltonian

\begin{align}\label{eq:heff}
    \hat{H} =& - \sum_{ ij, \sigma} t_{ij} z_{i} z_{j} c^{\dagger}_{i,\sigma}c_{j,\sigma}  - \mu \sum_{i,\sigma} c^{\dagger}_{i,\sigma}c_{i,\sigma}\\
    & + U  \sum_i \tilde{D}_i + \sum_{i,\sigma} V^{\text{imp}}_i c^{\dagger}_{i,\sigma}c_{i,\sigma} \nonumber\\
    & + \frac{V}{2} \sum_{i, j} \left[ \Delta^*_{ij} \left( c_{j, \downarrow}c_{i, \uparrow} + c_{i, \downarrow}c_{j, \uparrow} \right) \right. \nonumber\\
    & + \left.{\Delta_{ij}} \left(c^{\dagger}_{i, \uparrow}c^{\dagger}_{j, \downarrow} + c^{\dagger}_{j, \uparrow}c^{\dagger}_{i, \downarrow} \right) - 2 |\Delta_{ij}|^2 \right] \nonumber\\ 
       & - \frac{V}{2} \sum_{i,j} \left[ W^*_{ij} \left( c^{\dagger}_{j, \uparrow} c_{i, \uparrow} + c^{\dagger}_{j, \downarrow} c_{i, \downarrow}\right) \right. \nonumber\\
        &+ \left. W_{ij} \left( c^{\dagger}_{i, \uparrow} c_{j, \uparrow} + c^{\dagger}_{i, \downarrow} c_{j, \downarrow}\right) - 2 |W_{ij}|^2 \right] \nonumber\\
           & + \sum_i \left\{ \lambda_{1,i}  \left[ \hat{J}^{+}_i - \langle \hat{J}^{+}_i \rangle \right] + \lambda^*_{1,i}  \left[ \hat{J}^{-}_i - \langle \hat{J}^{-}_i \rangle \right] \right. \nonumber\\
    & \left. + \lambda_{2,i \sigma} \left( c^{\dagger}_{i,\sigma}c_{i,\sigma} - n_{i,\sigma} \right)  \right\} \;,\nonumber
\end{align}

with a mean-field decoupling of the nearest-neighbor interaction and

\begin{align*}
    n_{i,\sigma} &\equiv \langle c^{\dagger}_{i,\uparrow}c_{i,\uparrow} \rangle =  \langle c^{\dagger}_{i,\downarrow}c_{i,\downarrow} \rangle \;, \\
    \Delta_{ij} &\equiv \langle  c_{j,\downarrow} c_{i,\uparrow} \rangle  = \langle  c_{i, \downarrow} c_{j,\uparrow} \rangle  \;,\\
    W_{ij} &\equiv \langle c^{\dagger}_{j,\uparrow}  c_{i,\uparrow} \rangle  = \langle  c^{\dagger}_{j,\downarrow}  c_{i,\downarrow} \rangle  \,.
\end{align*}

The hamiltonian Eq. \ref{eq:heff} can be diagonalized via Bogoljubov-de Gennes
yielding the following eigenvalue equations

\begin{align*}
    \omega_{k} u_i(k) =& -\sum_{j} t_{ij} z_{i} z_{j} u_{j}(k) + \left\lbrack V^{\text{imp}}_i + \lambda_{3,i} -\mu\right\rbrack u_i(k) \\
        &- V \sum_{j}  W_{ij} u_{j}(k) + \lambda_{1,i} v_i(k)+  V\sum_{j}\Delta_{ij}v_{j}(k) \;,\\
    \omega_{k} v_i(k) =& \sum_{j} t_{ij} z_{i} z_{j} v_{j}(k) - \left\lbrack V^{\text{imp}}_i + \lambda_{3,i} -\mu\right\rbrack v_i(k) \\
        &+ V \sum_{j}  W_{ij} v_{j}(k) + \lambda_{1,i} u_i(k)+  V\sum_{j} \Delta_{ij}u_{j}(k) \;,
\end{align*}

which allows the calculation of the expectation values $n_{i,\sigma}$, $\Delta_{ij}$, $W_{ij}$, $\Delta_{i}=\langle   c_{i, \downarrow} c_{i,\uparrow}   \rangle$ and the total energy

\begin{align*}
    E\left(\left\{\Tilde{D}_i\right\}\right) =& - 2 \sum_{i, j} t_{ij} z^*_{i} z_{j} W^*_{ij} + U \sum_i \Tilde{D}_i \\
    & + V_1\sum_{i,j} (|\Delta_{ij}|^2 - |W_{ij}|^2)  + 2 \sum_i V^{\text{imp}}_i n_{i,\sigma} \;.
\end{align*}

This has to be minimized with respect to $\tilde{D}_i$. 
Finally, the Lagrange multipliers can be obtained from

\begin{align*}
    \lambda_{1,i} =& \frac{\partial E}{\partial \Delta^*_i} = - 4 \frac{\partial z_{i}}{\partial  \langle J_i \rangle } \frac{\partial \langle J_i \rangle }{\partial  \Delta^*_i} \sum_{j} t_{ij}  z_{j} \text{Re} \left(  W_{ij}\right) \;, \\ 
    \lambda_{2,i} =&  \frac{1}{2}\frac{\partial E}{\partial n_{i,\sigma}} = - 2 \frac{\partial z_{i}}{\partial  n_{i,\sigma}} \sum_{j} t_{ij}  z_{j} \text{Re} \left(  W_{ij}\right) \;.
\end{align*}

\bibliography{bibdata}

\end{document}